\documentclass[a4paper,11pt]{article}
\pdfoutput=1 % if your are submitting a pdflatex (i.e. if you have
\usepackage{jinstpub} % for details on the use of the package, please
\newcommand{\krm}{\isot{Kr}{83m}}
\newcommand{\Rb}{\isot{Rb}{83}}
\newcommand{\isot}[2]{$^{\textrm{#2}}$#1 }

\begin{document}
\title{\boldmath Analysis of \krm Prompt Scintillation Signals in the PIXeY Detector}

\author[a]{A.~G.~Singh,$^1$}\footnote{Now at: Stanford University, Department of Applied Physics, 348 Via Pueblo, Stanford, CA 94305}
\author[a]{A.~Biekert,}%
\author[a,b,c]{E.~Bernard,}%
\author[a,b,c]{E.~M.~Boulton,}%
\author[c]{S.~B.~Cahn,}%
\author[,d]{N.~Destefano,$^2$}\footnote{Now at: The MITRE Corporation, 202 Burlington Rd, Bedford, MA 01730}
\author[,c]{B.~N.~V.~Edwards,$^3$}\footnote{Now at: IBM Research, STFC Daresbury Laboratory, Warrington, WA4 4AD, UK}
\author[c,d]{M.~Gai,}
\author[,a,b,c]{M.~Horn,$^4$}\footnote{Now at: Sanford Underground Research Facility, 630 E. Summit Street, Lead, SD 57754}
\author[,c]{N.~Larsen,$^5$}\footnote{Now at: University of Chicago, Kavli Institute for Cosmological Physics, 5640 Ellis Ave, Chicago, IL 60637}
\author[,c]{B.~Tennyson,$^6$}\footnote{Now at: The MITRE Corporation, 202 Burlington Rd, Bedford, MA 01730}%%
\author[a,b]{Q.~Riffard,}%
\author[a]{V.~Velan,}%
\author[,c]{C.~Wahl,$^7$}\footnote{Now at: H3D, Inc., 3250 Plymouth Rd Suite 203, Ann Arbor, MI 48105}%
\author[a,b,c]{D.~N.~McKinsey}%
\affiliation[a]{University of California Berkeley, Department of Physics, Berkeley, CA 94720, USA}
\affiliation[b]{Lawrence Berkeley National Laboratory, Berkeley, CA 94720, USA}
\affiliation[c]{Yale University, Department of Physics, 217 Prospect St., New Haven, CT 06511, USA}
\affiliation[d]{University of Connecticut, LNS at Avery Point, 1084 Shennecossett Road, Groton, CT 06340, USA}

\emailAdd{agsingh@stanford.edu}

\abstract{Prompt scintillation signals from \krm calibration sources are a useful metric to calibrate the spatial variation of light collection efficiency and electric field magnitude of a two phase liquid-gas xenon time projection chamber.  Because \krm decays in two steps, there are two prompt scintillation pulses for each calibration event, denoted S1a and S1b.  We study the ratio of S1b to S1a signal sizes in the Particle Identification in Xenon at Yale (PIXeY) experiment and its dependence on the time separation between the two signals ($\Delta t$), notably its increase at low $\Delta t$.  In PIXeY data, the $\Delta t$ dependence of S1b/S1a is observed to exhibit two exponential components: one with a time constant of $0.05 \pm 0.02\mu s$, which can be attributed to processing effects and pulse overlap and one with a time constant of $10.2 \pm 2.2\mu s$ that increases in amplitude with electric drift field, the origin of which is not yet understood.}

\keywords{Cryogenic detectors, Dark Matter detectors, Interaction of radiation with matter, Noble liquid detectors, Time projection Chambers}

%\arxivnumber{1234.56789} % only if you have one

\maketitle
\flushbottom

\section{Background}
%------------------------------------------------------------
\subsection{Xenon Time Projection Chambers}
%------------------------------------------------------------
Two phase (liquid-gas) xenon time projection chambers (TPCs) have been used widely for dark matter detection.  When a particle interacts with the liquid xenon (LXe) in a TPC, this deposits energy which causes a burst of scintillation light and ionization of xenon atoms releasing free electrons, some of which recombine and produce additional light.  This light is detected by arrays of photomultiplier tubes (PMTs) on the top and bottom of the detector as the S1 signal.  The remaining free electrons are drifted upwards by applying an electric field and then extracted into the gaseous xenon (GXe) portion of the detector where they undergo electroluminescence producing a second signal, the S2 signal.  In general, the relative sizes of the S1 and S2 signals allows for discrimination between nuclear recoil and electron recoil events.  Additionally by examining which PMTs detected photons as well as the time delay between the S1 and S2 signals, 3D position reconstruction of the initial event is possible.  

\subsection{\krm as a Calibration Source}
Calibration of the detector's energy scale is critical.  \krm is effective for calibration, particularly useful as a mono-energetic standard candle for correcting any spatial non-uniformities in detector response.  It has a short half-life of approximately 1.83 hours and is able to dissolve into noble gases and distribute uniformly throughout a detector's volume. \krm also does not introduce any new long-lived radioisotopes into the detector.  For these reasons, \krm is a commonly used calibration source \cite{ks,lippincott,baudis,manalaysay,luxKr,chan,eckardt,decamp,stiller}.
%For these reasons, $^{83m}$Kr has been used in various two phase (liquid-gas) xenon time projection chambers (TPCs) utilized predominantly for dark matter detection

\begin{figure}
\begin{minipage}{0.46\textwidth}
\raggedright
\includegraphics[scale=0.274]{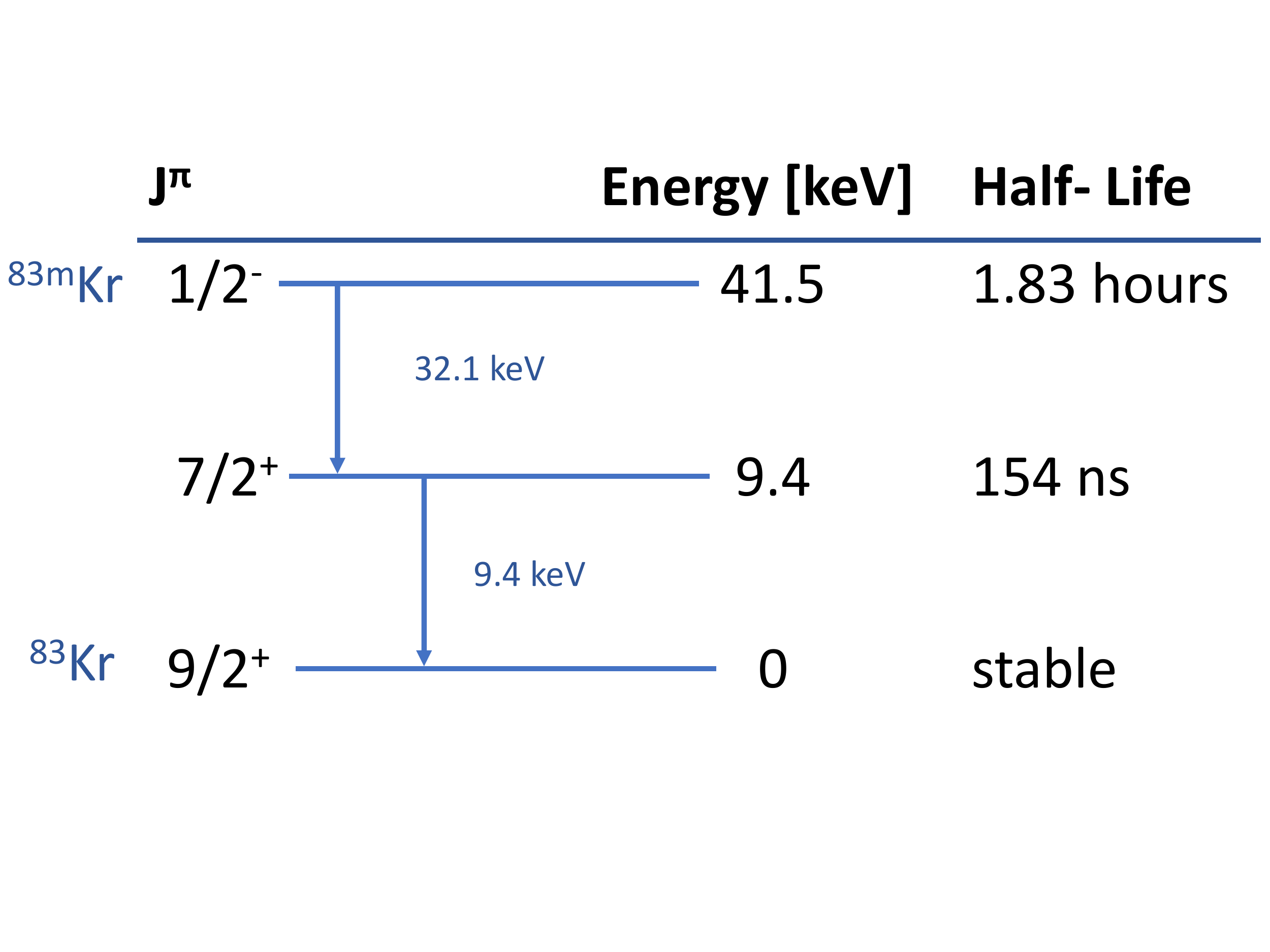}
\end{minipage}
\begin{minipage}{0.5\textwidth}
\footnotesize
%\raggedleft
%\captionsetup{type=table} %% tell latex to change to table
\begin{tabular}{||c| c |c||} 
 \hline\hline
 Transition & Decay Mode & Branching\\
 Energy & & Ratio[$\%$] \\[0.5ex] 
 \hline\hline
 32.1 keV & CE$_{M,N}$(32) & 11.5 \\ 
 & CE$_{L}$(30.4) + A(1.6) & 63.6\\
 & CE$_{K}$(17.8) + X$_{\kappa\alpha}$(12.6) + A(1.6) & 15.3\\
 & CE$_{K}$(17.8) + A(10.8) + 2A(1.6) & 9.4\\
 & $\gamma$ & <0.1\\
 \hline\hline
 9.4 keV & CE$_{L}$(7.5) + A(1.6) & 81.1 \\ 
 & CE$_{M}$(9.1) & 13.1\\
 & $\gamma$ & 5.8\\
 \hline\hline
\end{tabular}
\end{minipage}
\caption[Decay Sequence of $^{83m}$Kr]{Decay Sequence of $^{83m}$Kr: (a) Two-step transition (b) All possible de-excitation channels
\cite{branch1}\cite{branch2}}
\label{fig-decay}
\end{figure}

As demonstrated in Figure \ref{fig-decay}{\bf a}, $^{83m}$Kr decays to the ground state in two steps: first through a 32.1 keV transition and then through a 9.4 keV transition.  The possible de-excitation channels are shown in Figure \ref{fig-decay}{\bf{b}}.  For each of these transitions, the energy is primarily carried in internal conversion electrons (CE) or Auger electrons (A) and a smaller amount in x-rays (X$_{\kappa\alpha}$) and gamma rays ($\gamma$).  The resulting Compton scattering background interactions of the x-ray de-excitation channels are minimal, as photoabsorption dominates over Compton scattering at these energies.  Since $^{83m}$Kr decays in two steps there are two electrons, or incident particles, for each $^{83m}$Kr event.  Therefore, every $^{83m}$Kr event will have two prompt scintillation signals.  The prompt scintillation signal corresponding to the 32.1 keV emission is called the S1a signal while that corresponding to the 9.4 keV emission is called the S1b signal.  

In LXe, the electron-ion recombination fraction depends on both energy and electric field, and in the Large Underground Xenon experiment (LUX) it was found that the ratio of S1b to S1a amplitudes could be used to determine the magnitude of the drift field at any given location \cite{luxKr}.  This was particularly useful because any geometric effects on light collection should be the same for both S1a and S1b signals since they occur at the same location for any given event. Previous measurements of \krm scintillation in LXe have reported a dependence of the S1b scintillation signal on the time separation between the S1a and S1b signals \cite{delT,decay}. The S1 signals of \krm events provide a powerful calibration tool; however to be effective, the detector's response to these events must be well understood.  In this work we perform a detailed study of the S1b to S1a ratio in LXe for a variety of electric drift and extraction fields, using data acquired with the PIXeY detector.

%------------------------------------------------------------------
\subsection{Particle Identification in Xenon at Yale (PIXeY)}
\begin{figure}
\centering
\includegraphics[scale=0.4]{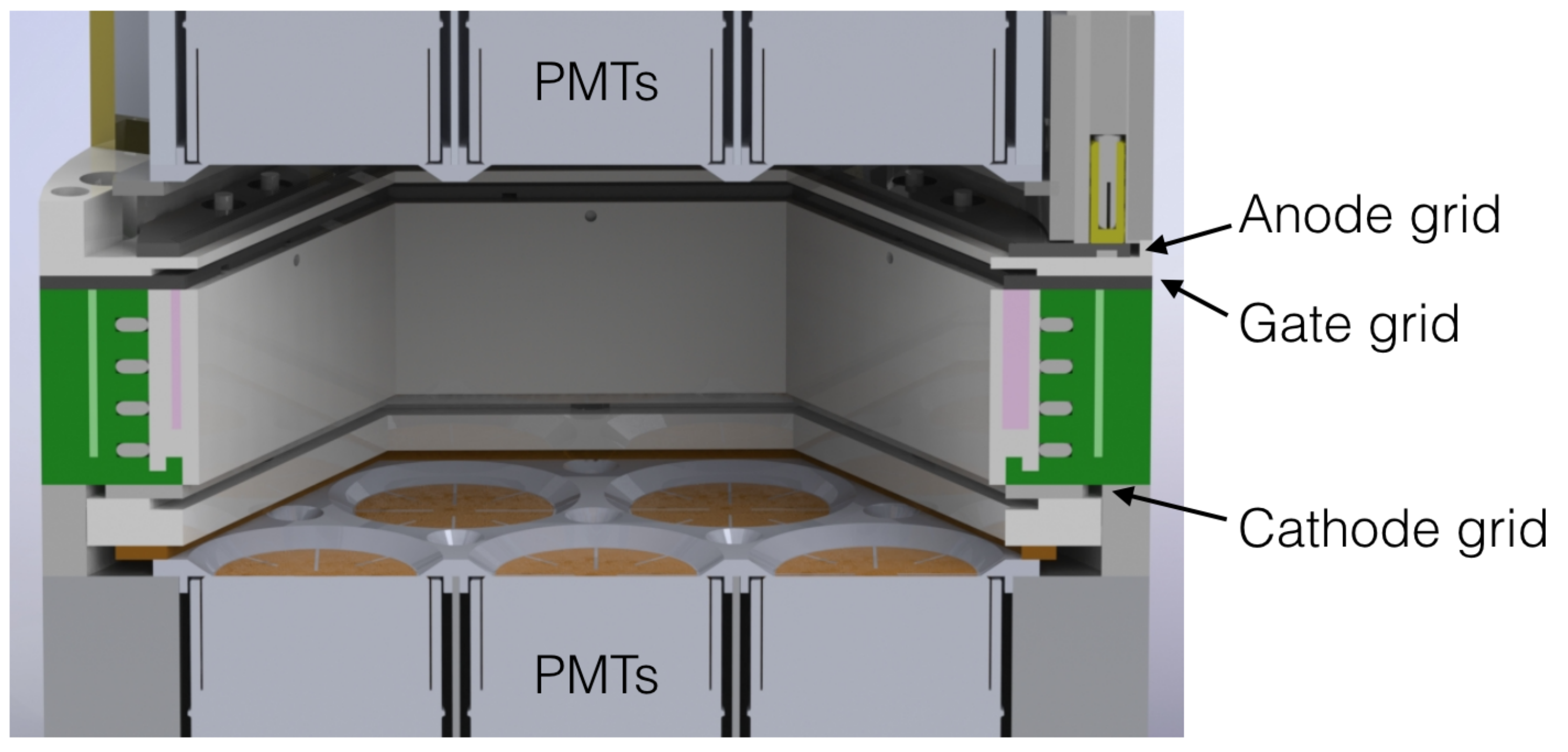}
\caption[Diagram of PIXeY:]{Cut-away rendering of PIXeY \cite{efficiency}.}
\label{fig:spec}
\end{figure}

%-------------------------------------------------------------------
PIXeY is a hexagonal two-phase xenon detector with a 5 cm drift length.  The rendering in Figure \ref{fig:spec} illustrates the components of the PIXeY detector.  At its widest point, the detector is 18.4 cm across and contains an active mass of 3 kg of xenon.  The gate lies 5.5 mm below the boundary between the liquid and gaseous portions of the detector \cite{pixey}.  The voltage difference between the cathode and gate grid and the anode and gate grid as noted set the drift field and extraction field respectively in the detector.  A wide range of voltages can be applied to the cathode to create drift fields between 50 and 2000 V/cm.

PIXeY has been used primarily to study the response of different radioisotopes such as $^{83m}$Kr, $^{37}$Ar, $^{137}$Cs, and $^{22}$Na in xenon time projection chambers as well as how that response varies as a function of drift and extraction fields \cite{pixey}\cite{d_thesis}.  These studies are crucial for characterization of electron recoil events \cite{firstLUX, 2017lux}.

%%%%%%%%%%%%%%%%%%%%%%%%%%%%%%%%%%%%%%%%%%%%%%%%%%%%%%%%%%%%%%%%%%%%%%%%%%%%%%%%%%%%%%

\section{\krm Events in PIXeY}

%%%%%%%%%%%%%%%%%%%%%%%%%%%%%%%%%%%%%%%%%%%%%%%%%%%%%%%%%%%%%%%%%%%%%%%%%%%%%%%%%%%%%%

%------------------------------------------------------------
\subsection{Synthesis and Injection of \krm into PIXeY}
%------------------------------------------------------------

The sample of \krm  was produced from \Rb from which it decays via pure electron capture.  The \Rb was synthesized by proton irradiation of a \Rb target performed at Brookhaven National Laboratory.  The resulting \Rb was then stored in an aqueous solution and diluted to achieve a desired activity level.  This \Rb solution was deposited on to a carbon mediator and baked under vacuum to bind and also remove water and volatiles.  The \Rb doped mediator was installed into the injection plumbing of the PIXeY detector.  The \krm generator was placed between a pressure differential in the GXe circulation path of the main chamber.  This pressure differential pushes the GXe over the mediator and into circulation.  The rate of GXe flow was dictated by the pressure differential produced by the main GXe circulation pump which is controlled  by a mass flow controller \cite{Kastens2}.

%------------------------------------------------------
\subsection{Data Acquisition and Processing}
%------------------------------------------------------

\begin{figure}[h]
\centering
\includegraphics[scale=0.457]{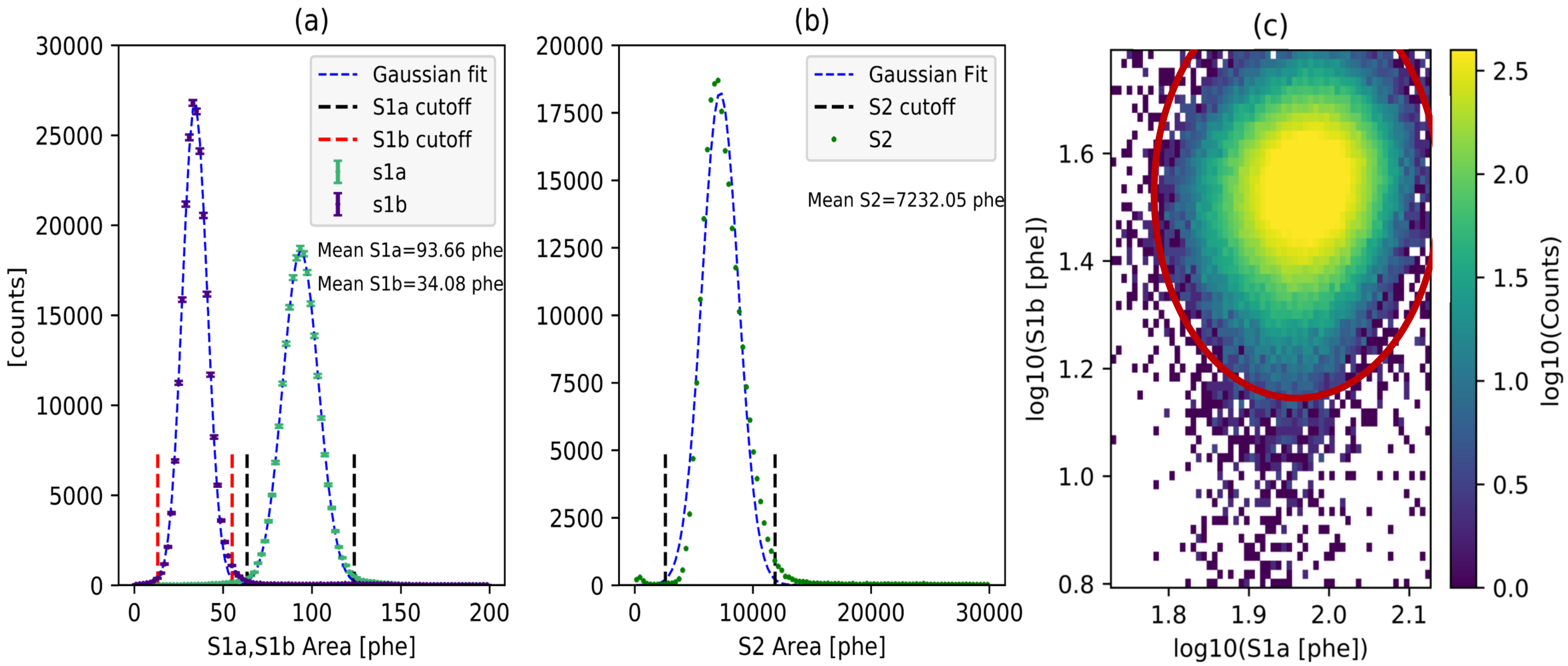}
\caption[One-Dimensional Fit of Signal Distributions]{{\bf (a)} Distributions of S1a and S1b signal sizes for 50V/cm data set {\bf (b)} Distribution of S2 signal size for 50V/cm data set {\bf (c)} 2D distribution of S1a and S1b detected signal sizes for 50V/cm data set.  Data outside of red line is cut.}
\label{fig:1dcut}
\end{figure}

Two arrays of seven Hamamatsu R8778 PMTs on the top and bottom of the detector were used to detect the \krm signals. The resulting data acquired from the PMTs underwent an eight-fold amplification before being digitized with a 12-bit ADC (CAEN V1720) at 250 MHz \cite{pixey}.  The digitized data was then processed to find events with two S1 pulses, one S2 pulse, and no additional pulses.  To ensure the selection of \krm events only, the remaining events were cut based on limits placed on S1a, S1b, and S2 size established by fitting each distributions of pulse sizes to a Gaussian distribution such as in Figure \ref{fig:1dcut}{\bf{a}} and {\bf{b}}.  Additional cuts were also implemented based on the fit of the 2-D distribution of the S1a signal size versus the S1b signal size as shown in Figure \ref{fig:1dcut}{\bf{c}}.  

\begin{figure}
\centering
\includegraphics[scale=0.6]{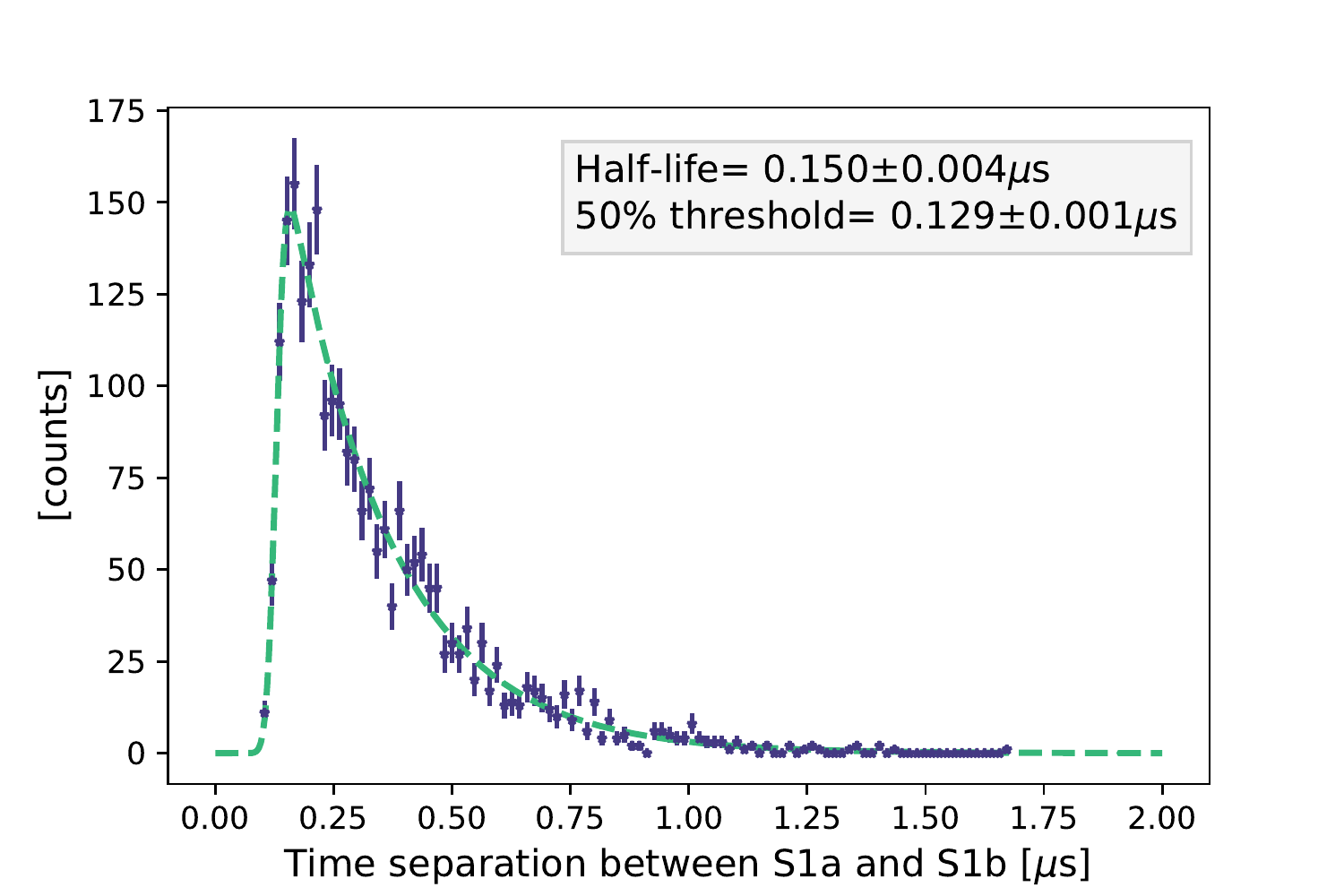}
\caption{Number of Double S1 Events Plotted as a Function of the Time Separation Between the S1a and S1b Signals: The distribution of signals is fit to the the product of the exponential decay and error function given by the dashed line.  From the returned fit, the half life value can be determined while additional parameters in the error function determine the low $\Delta t$ cut-off, labeled above as the 50$\%$ threshold, where the number of events detected with both S1 signals passes 50$\%$ of the maximum number of events observed.}
\label{eff}
\end{figure}

To avoid edge effects, events are also cut based on their location in the detector such that only a central fiducial volume of the detector is sampled.  The fiducial volume consists of a cylinder centered in the detector with a radius of 6.5 cm.  Additionally to reduce background from the PMTs, events with a drift time less than 5 microseconds or greater than 50 microseconds were cut.  This fiducial cut was applied to all drift fields studied.  Although the drift field will affect the maximum drift times of events, the overall values of S1b/S1a are observed to be relatively insensitive to this upper bound on drift time.  The purity of \krm event selection can be verified by plotting the number of events versus the time separation between the S1a and S1b signals in Figure \ref{eff}.  The shape of this distribution is consistent with the half life of the 9.4 keV transition of 154 ns.

\newpage

%%%%%%%%%%%%%%%%%%%%%%%%%%%%%%%%%%%%%%%%%%%%%%%%%%%%%%%%%%%%%%%%%%%%%%%%%%%%%%%%%%%%%%

\section{Description of Monte Carlo Simulation}

%%%%%%%%%%%%%%%%%%%%%%%%%%%%%%%%%%%%%%%%%%%%%%%%%%%%%%%%%%%%%%%%%%%%%%%%%%%%%%%%%%%%%%

%----------------------------------------------------
\subsection{Generating \krm Events in PIXeY}
%----------------------------------------------------

A Monte Carlo simulation of \krm events in PIXeY was developed to identify trends in the data that are effects of the detector's response or data processing.  To simulate a \krm event, first the number of photoelectrons in the event must be determined.  The mean number of photoelectrons in an S1a and S1b pulse for a given field was found by fitting distributions of S1a and S1b signal sizes from real data to a Gaussian distribution. From these mean values, a Poisson distribution was sampled to generate a value for the number of photoelectrons in a simulated pulse.  Based on the probability established by the singlet to triplet ratio, $f_s$, which has been experimentally determined \cite{hogenbirk}, a binomial distribution is sampled to determine the fraction of these photoelectrons assigned to be in the singlet portion of the pulse while the remaining photoelectrons are assigned to the triplet portion of the pulse.  These photoelectrons are then distributed over some period of time via the sampling of the following exponential distribution 

\begin{gather}\label{eq:1}
I(t)= f_s(\frac{1}{\tau_s} e^{\frac{-t}{\tau_s}})+(1-f_s)(\frac{1}{\tau_t} e^{\frac{-t}{\tau_t}})\tag{1}
\end{gather}

\noindent where $\tau_s$ and $\tau_t$ are the time constants of the singlet and triplet pulses, determined to be 3.1 ns and 24 ns respectively \cite{lifetime_original}\cite{discriminate}.

\begin{figure}[h]
\centering
\includegraphics[scale=0.375]{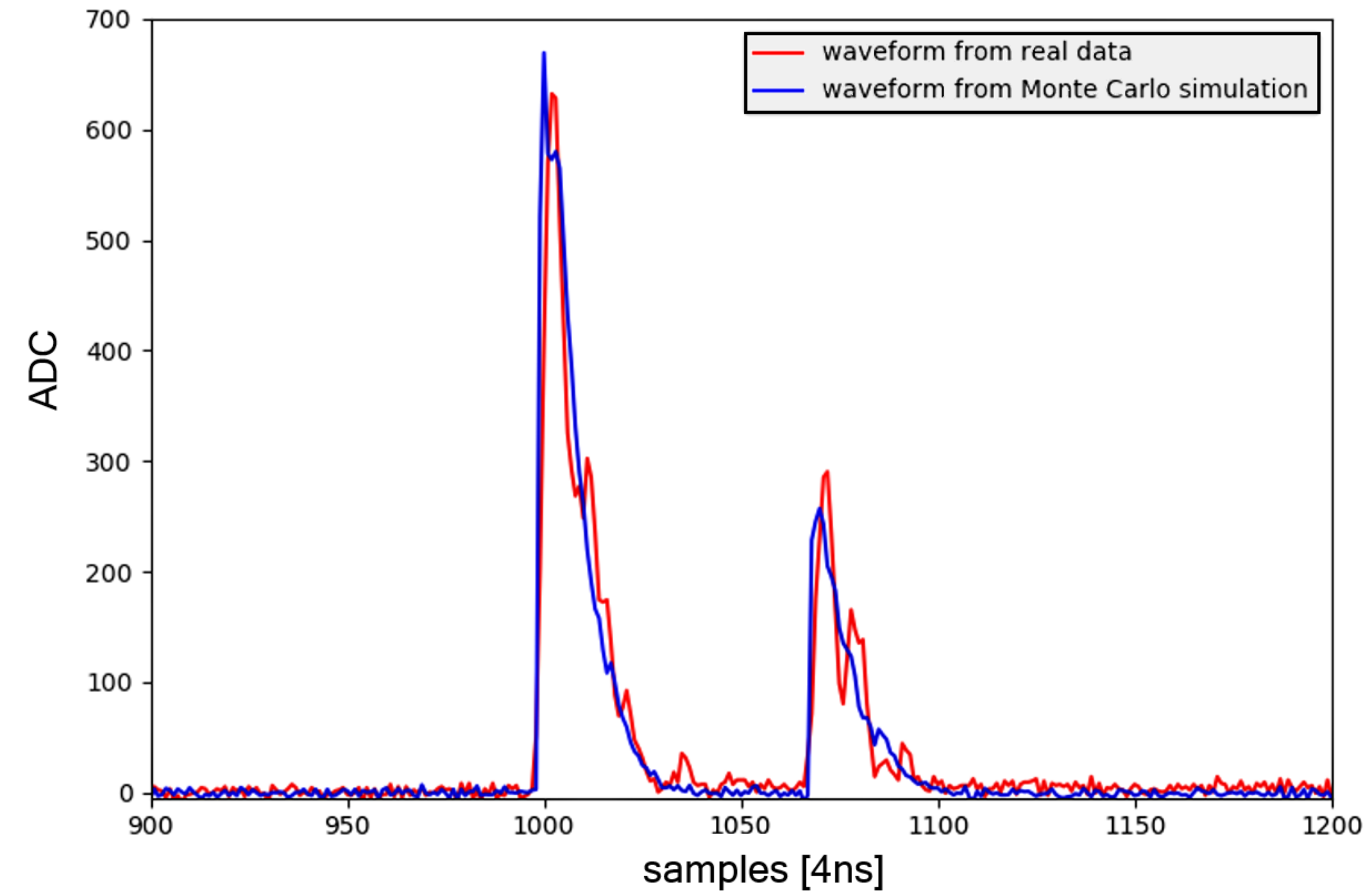}
\caption{Overlay of a simulated and real wave form event from data.  Simulated event was constructed with the same value of photoelectrons in each pulse and start times}
\label{compare}
\end{figure}

This distribution of pulses was then transformed by a series of procedures to produce wave forms.  The time scale was re-scaled to replicate the sampling rate of PIXeY.  To generate a wave form for each of the PIXeY PMT channels, the event pulses were split under the assumption that each photoelectron had an equal probability of being detected by any of the PMTs. The magnitude of the pulses were then converted from number of photoelectrons to ADC using a conversion based on the single electron size found in the data.  Each channel's wave form was then convolved with a PMT transfer function consistent with what we observe in PIXeY.  The wave forms were then offset, inverted and Gaussian white noise was added to the simulated signal to match what a real PIXeY event would like in the data. In Figure \ref{compare}, the simulated  and real wave forms are overlaid, demonstrating general agreement between the two.

%%%%%%%%%%%%%%%%%%%%%%%%%%%%%%%%%%%%%%%%%%%%%%%%%%%%%%%%%%%%%%%%%%%%%%%%%%%%%%%%%%%%%%

\section{Time Separation between S1 Signals and Effect on Signal Size}

%%%%%%%%%%%%%%%%%%%%%%%%%%%%%%%%%%%%%%%%%%%%%%%%%%%%%%%%%%%%%%%%%%%%%%%%%%%%%%%%%%%%%%

%--------------------------------------------------------------
\subsection{Enhanced S1b/S1a at Low $\Delta t$}
%--------------------------------------------------------------

\begin{figure}[h]
\centering
\includegraphics[scale=0.55]{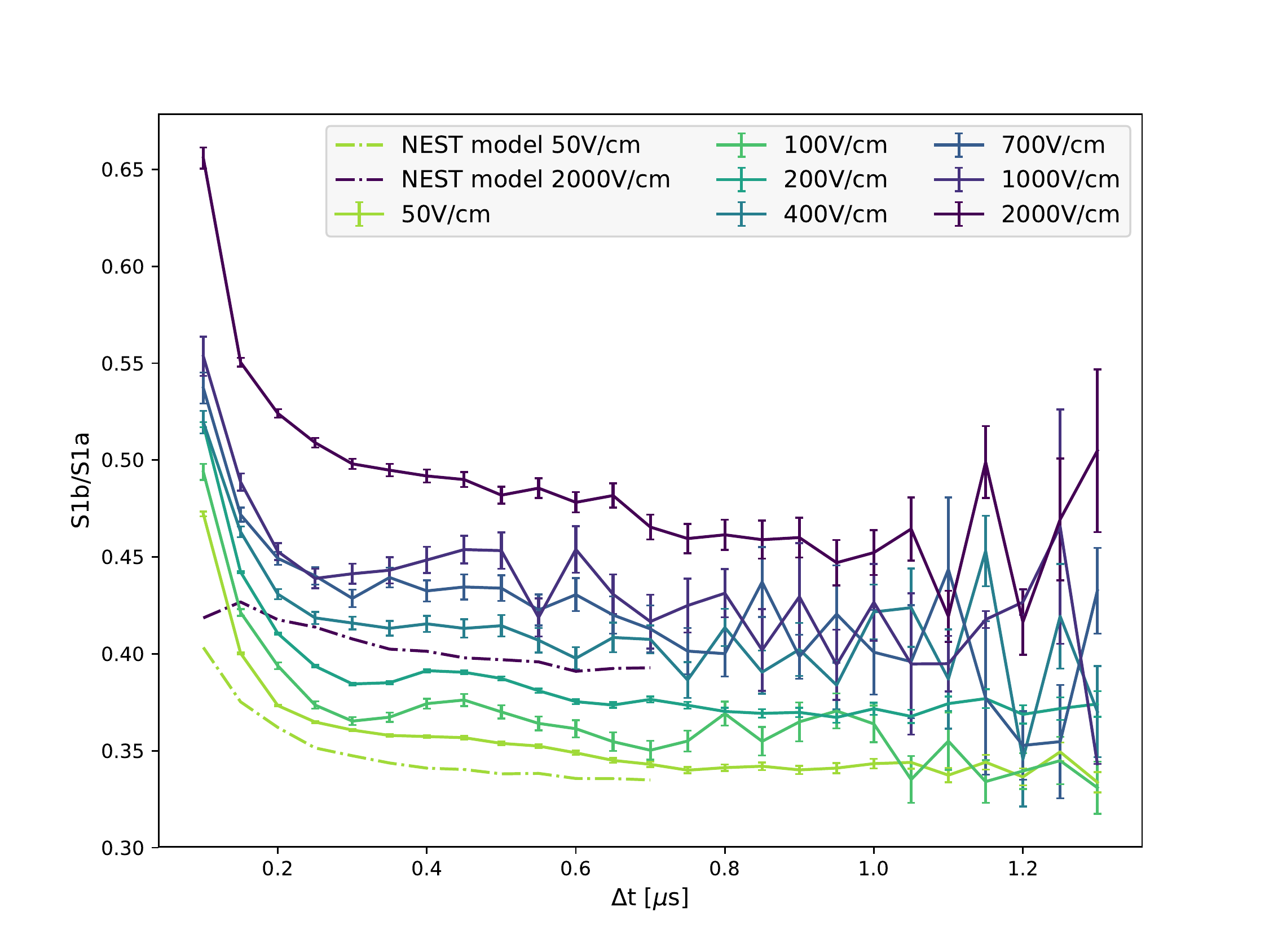}
\caption{{\bf Average S1b/S1a as a Function of $\Delta t$:} the ratio of S1b signal size to S1a signal size increases as the time between the two signals decreases.  This effect is demonstrated over all applied fields for which data was collected.  The data is also plotted with NEST simulations at 50 and 2000V/cm. \cite{nest} It is notable that disagreement between the model and data grows with field and also at lower $\Delta t$.  At 50V/cm, PIXeY data disagrees with NEST simulations by 2-4\% and grows to 6\% at low $\Delta t$, while disagreement at 2000V/cm, ranges between 15\% and 20\% and grows to as much as 29\% at the lowest values of $\Delta t$.}
\label{s1b}
\end{figure}

The time difference between S1a and S1b signals is denoted $\Delta t$. An increase in the size of the S1b signal size at low $\Delta t$ has been consistently reported \cite{delT} \cite{decay} and is also observed here.   The PIXeY data sets also demonstrate a decrease in S1a signal size at low $\Delta t$. On account of both of these effects the value of S1b/S1a has a strong dependence on $\Delta t$ as demonstrated in Figure \ref{s1b} across all drift fields.  There are several hypotheses regarding the origin of this effect. Slight overlap of the S1a and S1b scintillation pulses can result in S1a photons falling within the S1b pulse, thus decreasing the reconstructed S1a and increasing S1b.  Diffusion of left-over charges from the S1a signal might also affect recombination in the S1b pulse \cite{delT}.  The simulated \krm events can be used to investigate these hypotheses.  The number of photoelectrons in the simulated pulses is sampled from a Poisson distribution with a rate set by the average S1a and S1b signal sizes at high $\Delta t$ where no trends are observed. 

%--------------------------------------------------------------------
\subsection{Comparison Between Monte Carlo Simulation and Real \krm Events}
%--------------------------------------------------------------------

\begin{figure}[h]
\centering
\includegraphics[scale=0.55]{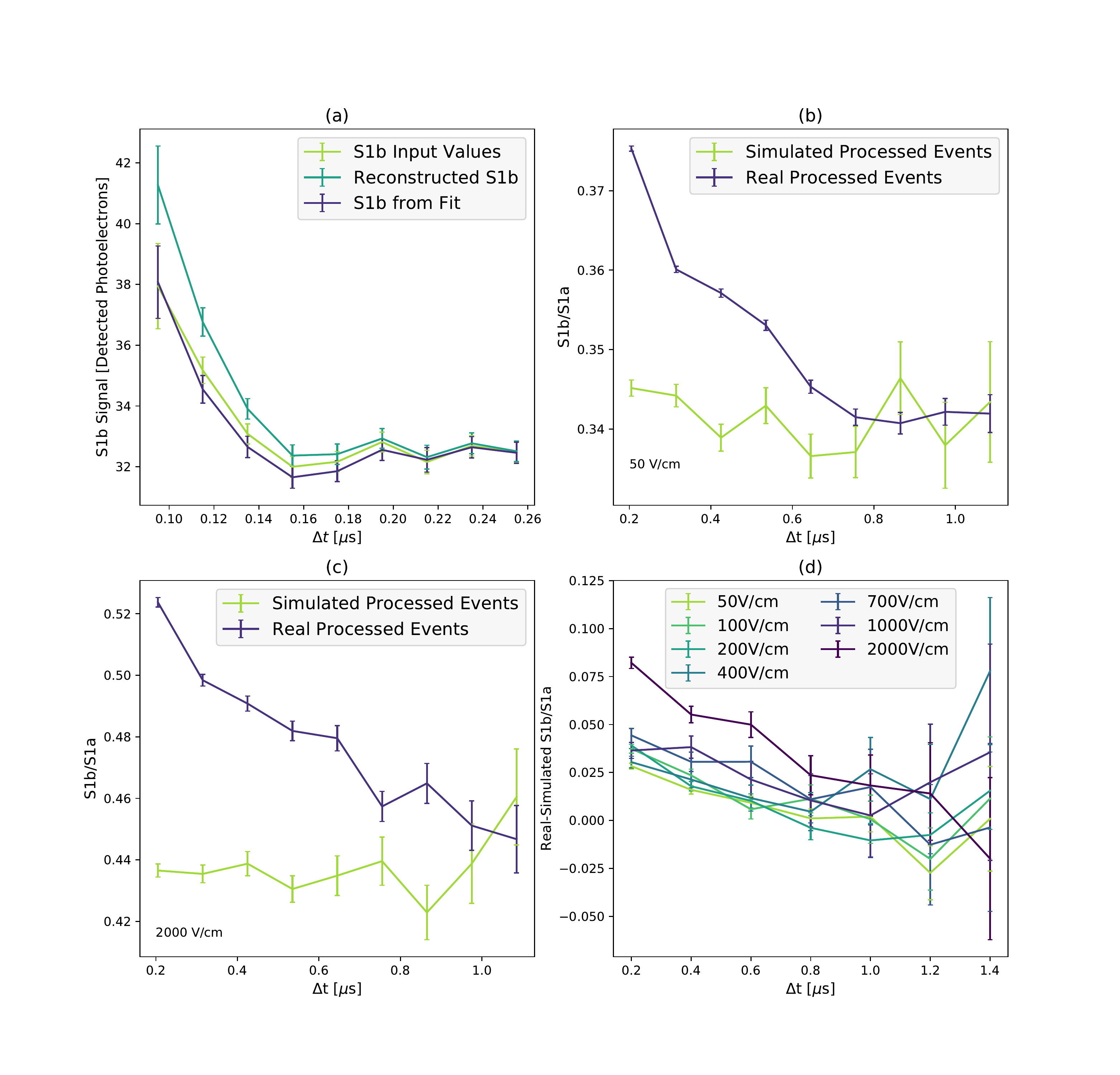}
\caption{{\bf a)} Different measurements of S1b/S1a for simulated events at 50V/cm {\bf b-c)} Average S1b/S1a as a function of $\Delta t$ calculated for real and simulated data taken at 50 and 2000 V/cm respectively. {\bf d)} Non-zero difference between experimental and simulated S1b/S1a ratio at low $\Delta t$}
\label{fig7}
\end{figure}

Samples of simulated events were developed with a $\Delta t$ value set by sampling a time distribution based on the half-life of $^{83m}$Kr.  The resulting sample wave forms then underwent the same processing that was conducted on the real wave form data.  Some examples of the simulation results are pictured in Figure \ref{fig7}.  The simulated data also present some increase in S1b/S1a size at low $\Delta t$ on an account of an increase in the size of the S1b signal and a decrease in the S1a signal size.  However, overall, the trends in the real and simulated values are not consistent. 

For small values of $\Delta t$ it is feasible for the tail of the S1a pulse to overlap with the beginning of the S1b pulse given the triplet lifetime is significant on this timescale.  Additionally the response of the PMTs and electronics add some finite time response to the signal.  Simulated wave forms at low values of $\Delta t$ can recreate this pulse overlap effect.  As seen in the plots of the simulated wave forms in Figure \ref{fig7}{\bf a}, a sharp increase in the size of the S1b signal is observed for $\Delta t$ less than 200ns.

To separate possible physical trends in these data from processing artifacts, we can modify how we assess the size of the pulses.  Rather than integrating the raw wave forms, the S1a, S1b event wave forms can be fit to the sum of two exponential functions.  From the fit returned, the functional form of each exponential peak can be extrapolated.  These individual peaks can then be integrated to determine the area of the pulses now with any overlap effects accounted for and eliminated.

In Figure \ref{fig7}{\bf a}, the results of fitting simulated wave forms are shown.  The simulated wave forms provide a powerful tool of comparison since the original event size input into the simulation is known.  These input values as a function of $\Delta t$ are plotted in green in Figure \ref{fig7}{\bf a}.  These values are drawn randomly from a Poisson distribution and therefore should have no dependence on $\Delta t$.  The data plotted however demonstrates an upturn at low $\Delta t$.  This demonstrates a bias in the event discrimination of the data processing.  For accurate detection of double S1 events at low $\Delta t$ to occur, the relative size of the S1b signal must be larger.  This processing effect creates a baseline shift in the size of the S1b signals observed in PIXeY data. 

In addition, the data also display a divergence between the size of the S1b signal returned from fitting the wave forms and that returned from the processing code, the reconstructed quantity (RQ).  As expected, the values returned from the fit are consistent with the true values input into the simulation.  The increase in RQ values at low $\Delta t$ then can be attributed to pulse overlap.  From Figure \ref{fig7}{\bf a}, the effective time scale for this effect is for $\Delta t$ < 200 nanoseconds.  In line with this, the decrease in S1a signal size in the data also occurs on this same timescale.

In addition to this sharp increase in S1b/S1a below 200 nanoseconds, real PIXeY events also demonstrate a shallow feature in the data above 200 nanoseconds that occurs on a significantly longer time frame,  as illustrated in Figure \ref{fig7}{\bf b-d}. In Figure \ref{fig7}{\bf b-c}, while the average S1b/S1a for simulated events approaches a constant value for $\Delta t$ greater than 200 ns, in the experimental data there is an additional dependence on $\Delta t$.  In Figure \ref{fig7}{\bf d}, the difference between the average simulated S1b/S1a and the experimentally determined S1b/S1a is plotted as a function of $\Delta t$.  By subtracting the simulated ratio values, all effects due to processing should be canceled out.  The difference plotted however shows a residual trend with $\Delta t$ that also appears to increase with drift field.

There are several systematic effects in PIXeY that could artificially cause this slow rise in S1b/S1a at low $\Delta t$.  PMT ringing, for example, could in principle lead to enhanced S1b signal size.  To explore this effect, an averaged event wave form can be fit to the same exponential model previously used and the average residual of the S1b pulse from this fit can be calculated.  If PMT ring down was systematically affecting the data, these residuals would be expected to follow a trend with $\Delta t$.  However, these residuals show no dependence on $\Delta t$ making it unlikely that PMT ring down has a significant systematic effect on the apparent size of the S1b signal. If the S1a pulse produced a time-dependent PMT gain over the next few hundred nanoseconds, this could in principle create a $\Delta t$-dependent S1b/S1a ratio; however the PIXeY PMT voltage dividers are designed for excellent linearity in this pulse size regime. Furthermore, this slow decay increases in magnitude as electric field increases, while S1a decreases with increasing drift field, so a PMT gain effect would have the opposite field dependence to what is observed.  If the electron lifetime in the detector is correlated with the applied electric field, this could possibly enhance the slow component of the S1b pulse.  Over all field values however, the electron lifetime is much greater than the maximum electron drift time in the detector and is not correlated with electric field.  It is possible then that the slow trend observed in S1b/S1a may be a physical consequence of change to recombination behavior in the detector.

One suggested model for the $\Delta t$ -dependence proposes an enhancement in S1b recombination due to left-over electron-ion pairs from the S1a event that failed to recombine \cite{delT}.  The increase in S1b signal size as a function of $\Delta t$ would therefore be proportional to the diffusion of this cloud of charge as a function of time.  The density of this cloud can be modeled as $ n(t) \propto (2t+\tau_D)^{-3/2}$ where $\tau_D$ is the diffusion timescale defined as $\frac{a^2}{D}$.  $D$ is the diffusion coefficient of electrons in LXe while $a$ is the characteristic length scale of the initial distribution, here determined by the typical range of a 30 keV electron in LXe. Using these values, $\tau_D$ is estimated in \cite{delT} to be 10 ns in the case of zero field. When a field is applied, there will be less of a diffusion effect since the applied field provides an additional way for electrons to leave the area of the event.
%In \ref{fig6_rev}, this diffusion model is plotted by the dashed red line with the value of $\tau_D$ suggested by the discussion outlined in \cite{delT} for zero field.  These models suggest a feature with a timescale on the order of 10ns.  While the PIXeY data also has a feature on this timescale, the model does not reproduce the additional slow feature in the data . Consequently this model for enhanced recombination can not be explicitly verified.

%\begin{figure}[h]
%\centering
%\includegraphics[scale=0.7]{s1b_zoom.png}
%\caption{Diffusion model with time constants modified for applied electric field \cite{baudis} overlaid with real PIXeY data.  Overall, these models suggest a feature with a shorter time scale compared to the slope we see in the real data.}
%\label{s1b_zoom}
%\end{figure}

%%%%%%%%%%%%%%%%%%%%%%%%%%%%%%%%%%%%%%%%%%%%%%%%%%%%%%%%%%%%%%%%%%%%%%%%%%%%%%%%%%%%%%%
\section{S1b/S1a in PIXeY}
%%%%%%%%%%%%%%%%%%%%%%%%%%%%%%%%%%%%%%%%%%%%%%%%%%%%%%%%%%%%%%%%%%%%%%%%%%%%%%%%%%%%%%%

\subsection{Developing a Model for S1b/S1a}

%\begin{figure}[h]
%\centering
%\includegraphics[scale=0.9]{spatial_dist_vs_ratio100.png}
%\caption{{\bf Size of S1b/S1a signal ratio is independent of event location in detector} The spatial distribution of events in the detector is the same for various ranges of ratio sizes}
%\label{spat_dist}
%\end{figure}

The ratio of S1b to S1a signal size is a useful metric to study since it facilitates comparison between detectors, as detector-specific behavior  and corrections will cancel out, and moreover it provides insight into different physical behavior in LXe \cite{Farham}.  As previously discussed, since S1b/S1a has a clear dependence on the applied field, the local value of S1b/S1a has been used as a measurement to assess fluctuations in the electric field throughout the detector volume \cite{luxKr}.  PIXeY has a comparatively small active volume, and minimal variation in the actual electric field within the detector would be expected.  %Supporting this claim, as demonstrated in \ref{spat_dist}
Consequently in PIXeY the size of the S1 signal has little dependence on the location of the event in the detector.  However, as demonstrated in the previous section, the ratio value has a strong dependence on the time between the S1a and S1b pulses.

As a result the value of this ratio is highly dependent on the detector's specific ability to discriminate double S1 events at low $\Delta t$.  If a detector has poor detection efficiency at low $\Delta t$ or these events are otherwise cut during data processing this can artificially decrease the average value of S1b/S1a.  To that end, a developed understanding between the relationship between $\Delta t$ and S1 signal size is crucial to characterize the behavior of the detector and accurately interpret the signal data produced.

An empirical model for S1b/S1a in PIXeY can be developed where this dependence on $\Delta t$ is incorporated, as well as the dependence on the applied electric field.  The data in Figure \ref{s1b} were fit to a model based on the sum of two exponential decays:

\begin{gather}\label{eq:2}
\frac{S1b}{S1a}(\Delta t)= a*\exp{\Big{(}-\frac{\Delta t}{b}}\Big{)}+c*\exp{\Big{(}-\frac{\Delta t}{d}}\Big{)} \tag{2}
\end{gather}

\begin{figure}
\includegraphics[scale=0.481]{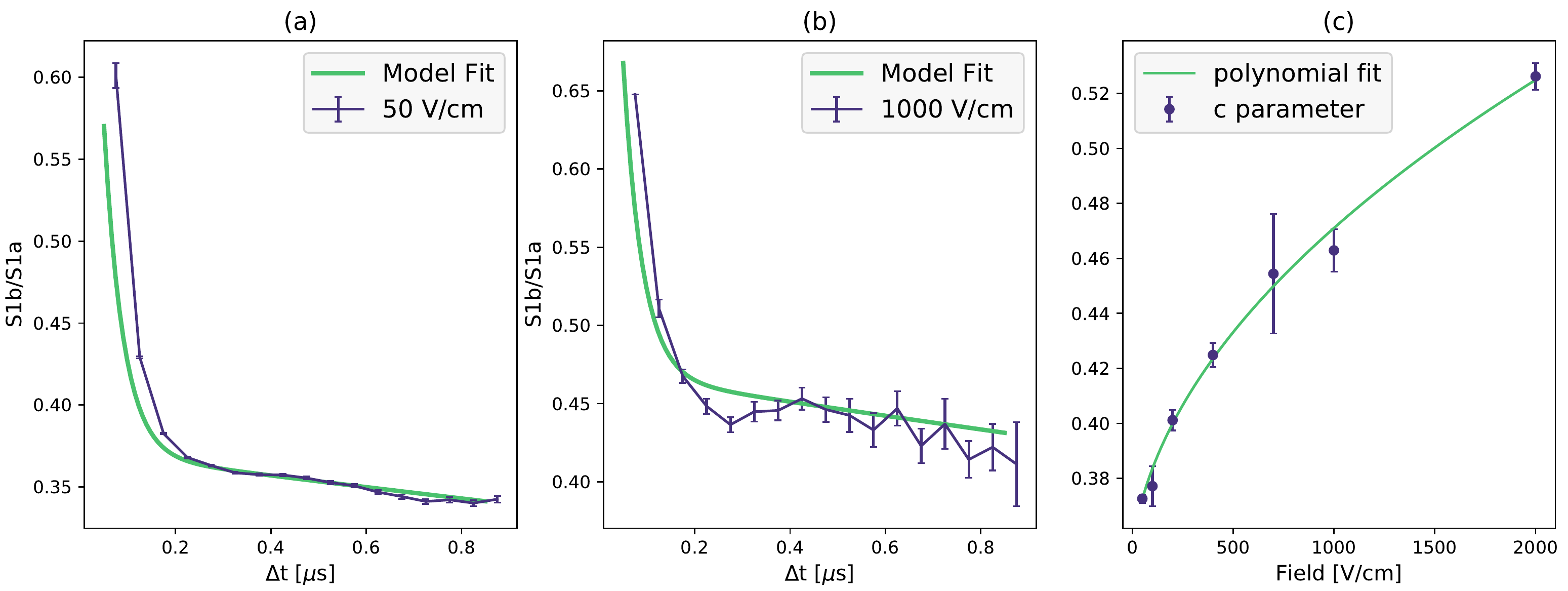}
\caption{{\bf a-b)} Empirical model for S1b/S1a compared to two data sets {\bf c)} Dependence of c coefficient in model on applied electric field}
\label{model}
\end{figure}

Across all fields, the parameters $a$, $b$ and $d$ were effectively constant and were found to be a= 0.11+/-0.03, b= 0.05+/-0.02 $\mu s$ and d= 10.2+/-2.2 $\mu s$.  The values for the coefficient $c$, which gives the amplitude of the slow component, has a clear dependence on the applied field as demonstrated in Figure \ref{model}{\bf c}.  This field dependence can be modeled well by a polynomial.  We find: $$c=0.34+0.02\sqrt{0.06E+0.53}$$ where $E$ is the electric drift field in units of V/cm.  %NEST simulations of \krm events find the fraction of electrons from the S1a pulse that fail recombine follow a similar dependence on drift field as the amplitude coefficient, c.  This could support that enhanced recombination due to leftover charges from the S1a signal contribute to the slow component in the S1b/S1a trend with $\Delta t$.  
The model was found to recreate S1b/S1a data effectively as demonstrated in Figure \ref{model}{\bf a-b}.

\begin{figure}[h]
\centering
\includegraphics[scale=0.5]{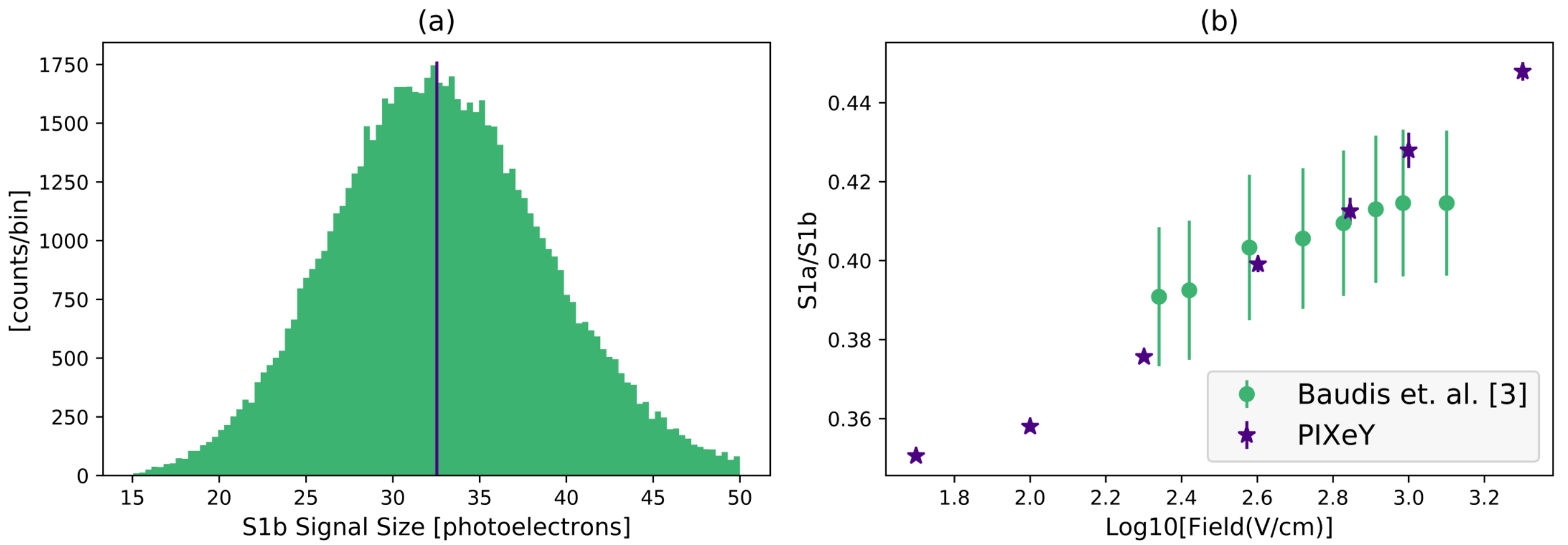}
\caption{{\bf a)} Distribution of S1b signals for a data set.  Purple bar indicates the mean returned from fit of distribution to Gaussian {\bf b)} Ratio as a function of applied field for PIXeY compared to \cite{baudis}: error bars refer to standard error on both means propagated.  In each of these figures, for PIXeY data, only events with a $\Delta t$ value greater than 300 ns were included in the analysis, effectively cutting the events with the most dramatic dependence on $\Delta t$.}
\label{ratio}
\end{figure}

If instead the value of $\Delta t$ is held constant by only selecting events which have a value of $\Delta t$ that fits within a fixed window, the average value of S1b/S1a can be studied simply as a function of applied electric field. Controlling for the $\Delta t$ dependence on the ratio value is necessary if the ratio value is to be used to compare between other detectors.  The average ratio for each field level was calculated using a peak fitting method.  For each field data set a histogram like that shown in Figure \ref{ratio}{\bf{a}} was created of all the S1a and S1b data.  This histogram was fit to a Gaussian distribution and the mean of the fit was returned.  The S1a and S1b means found in this manner are divided to estimate the mean ratio value for that field.  The mean ratio values calculated for each field level in PIXeY are presented in Figure \ref{ratio}{\bf{b}} along with corresponding values presented in \cite{baudis}.  With ratio variation with $\Delta$t in PIXeY now controlled for, overall the ratios found in both detectors are consistent over the range of fields for which data is reported. 

%%%%%%%%%%%%%%%%%%%%%%%%%%%%%%%%%%%%%%%%%%%%%%%%%%%%%%%%%%%%%%%%%%%%%%%%%%%%%%%%%%%%%%

\section{Conclusion}

Using the PIXeY detector, a detailed study of the S1b to S1a signal ratio in \krm events was performed for a variety of drift fields, showing a $\Delta t$ dependence of S1b/S1a with two exponential components. The fast component is consistent with expectations from scintillation pulse overlap, while the slower component is unexplained and increases with the magnitude of the drift field. Systematic effects from detector behavior that could create this trend have been largely ruled out. This slow component might then be due to recombination effects as proposed in \cite{delT,decay}.  The model proposed in \cite{delT} has a power law form, while here we find good agreement with an exponential decay. Reference \cite{decay} reports data acquired at zero electric drift field, and shows a larger percentage decrease in the size of the S1b scintillation signal with increasing $\Delta t$.    

%%%%%%%%%%%%%%%%%%%%%%%%%%%%%%%%%%%%%%%%%%%%%%%%%%%%%%%%%%%%%%%%%%%%%%%%%%%%%%%%%%%%%%

\acknowledgments

We acknowledge support from DHS grant 2011-DN-007-ARI056-02, NSF grant PHY-1312561, and DOE grant DE-FG02-94ER40870.  The $^{83}$Rb used in this research to produce \krm was supplied by the United States Department of Energy Office of Science by the Isotope Program in the Office of Nuclear Physics.


%%%%%%%%%%%%%%%%%%%%%%%%%%%%%%%%%%%%%%%%%%%%%%%%%%%%%%%%%%%%%%%%%%%%%%%%%%%%%%%%%%%%%%%%%%%%

% Please avoid comments such as "For a review'', "For some examples",
% "and references therein" or move them in the text. In general,
% please leave only references in the bibliography and move all
% accessory text in footnotes.

% Also, please have only one work for each \bibitem.

\end{document}